\documentclass[sigconf]{acmart}

\setcopyright{acmcopyright}
\copyrightyear{2022}
\acmYear{2022}

\acmConference[CSLAW '22]{ACM Symposium on Computer Science and Law}{November 01--02, 2022}{Washington, DC}
\acmBooktitle{CSLAW '22: ACM Symposium on Computer Science and Law, November 01--02, 2022, Washington, DC}

\begin{document}

\title{Programming Languages and Law}
\subtitle{A Research Agenda}

\author{James Grimmelmann}
\email{james.grimmelmann@cornell.edu}
\orcid{0000-0002-9330-7698}
\affiliation{
  \institution{Cornell University}\department{Law School and Cornell Tech}
   \city{New York City}
   \state{NY}
   \country{USA}
}

\begin{abstract}
 If code is law, then the language of law is a programming language. Lawyers and legal scholars can learn about law by studying programming-language theory, and programming-language tools can be usefully applied to legal problems. This article surveys the history of research on programming languages and law and presents ten promising avenues for future efforts. Its goals are to explain how the combination of programming languages and law is distinctive within the broader field of computer science and law, and to demonstrate with concrete examples the remarkable power of programming-language concepts in this new domain.
 \end{abstract}



\maketitle

\section{Introduction}

Computer science contains multitudes. It ranges from pure mathematics to quantum physics, from the heights of theory to the depths of systems engineering. 

Some of its subfields speak to urgent problems law faces. Criminal procedure\cite{kerr2018computer} and national security law\cite{chesney2021cybersecurity} cannot regulate the world as it exists without taking account of whether, when, and how data can be kept private. Other subfields provide new perspectives on law. The "law as data" movement\cite{ashley2017artificial,livermore2019law} uses computational methods like topic modeling and decision-tree learning to analyze legal datasets in subejcts as diverse as trademark infringement,\cite{beebe2006empirical} judicial rhetoric,\cite{livermore2017supreme} and the structure of the United States Code.\cite{katz2014measuring}

I would like to argue that the computer-science field of \emph{programming-language theory} (PL theory) speaks to law in both of these senses. Not only is it indispensable for answering certain kinds of practical legal questions, but its application can "illuminate the entire law."\cite{easterbrook1996cyberspace} Not every part of computer science speaks to law in this way, but PL theory does.

That is, PL theory can be the $x$ in defining a meaningful subfield of "law and $x$," in which in which the scholarly methods of $x$ provide novel ways to formulate and answer legal questions. Just as microeconomics provides a new and illuminating way to think about tort remedies, and just as corpus linguistics provides a new and illuminating way to think about statutory interpretation, the PL theory provides new and illuminating ways to think about familiar issues from all across the law.

Consider, for example, the M++ project to formalize French tax law (described in more detail in Section~\ref{sec:frenchtax}), which is helping the French tax authority upgrade its internal computer systems. This by itself is not particularly remarkable; tax authorities around the world have systems to compute taxpayers' obligations. Instead, M++ is distinguished by its rigorous use of PL theory to design a new programming language for describing the provisions of the French tax code. On the one hand, M++ is useful because it is a clean, modern language that is amenable to correctness proofs, improving the reliability of tax computations. On the other hand, M++ attempts to mirror the structure of the tax laws it formalizes. Instead of treating the rules of tax law as an \emph{ad hoc} design document, M++ treats the tax code \emph{as though they were itself a program}, one meant to be "executed" by lawyers and accountants. The goal is not just to do the same thing as the tax code, but to do it in the same way, section by section, clause by clause.

To generalize, PL theory has something unique to offer law because there is a
crucial similarity between lawyers and programmers: the way they use
\emph{words}. Computer science and law are both \emph{linguistic} professions,
whose practitioners use language to create, manipulate, and interpret complex abstractions. A programmer who uses the right words in the right way makes a computer do something. A lawyer who uses the right words in the right way changes people’s rights and obligations. There is a nearly exact analogy between the text of a program and the text of a law.

This parallel creates a unique opportunity for PL theory as a
discipline to contribute to law. Some CS subfields, such as AI, deal with legal
structures. Others, such as natural language processing (NLP), deal with legal
language. But only PL theory  provides a principled, systematic
framework to analyze legal structures in terms of the linguistic expressions
lawyers use to create them. Programming-language abstractions have an unmatched
\emph{expressive} power in capturing the linguistic abstractions of law.

Over a decade ago, Paul Ohm proposed a new research agenda for "computer programming and law," describing in detail the value of executable code for legal scholarship: by gathering and analyzing information about the law more efficiently, by communicating more effectively, and by writing programs to test and prove scholarly points.\cite{ohm2009computer} More recently, Houman Shadab also called for legal scholars to improve their research by learning to code.\cite{shadab2020software} In one sense, this article takes up their call: PL theory can help legal scholars write better code. In another, it makes a more ambitious claim: if code is law,\cite{lessig1999code} then \emph{understanding code is a way of understanding law, and vice versa}. PL theorists understand code in a profound and distinctive way; legal scholars understand law in a profound and distinctive way. They can and should learn from each other. 

Indeed, they already are. Parts of this article are forward-looking, describing the possibilities for what PL theory and law could become. But much of the work of laying out a research agenda is simply looking back and taking stock of what has already been accomplished. There is an active community of PL theory and law researchers, drawing on experts from both sides and building collaborations between the two. In January 2022, the ACM held the first Programming Languages and Law workshop (ProLaLa) as part of the annual ACM SIGPLAN Symposium on Principles of Programming Languages (POPL). Much of the work I will describe was presented there; indeed, I gave an earlier version of this article as a ProLaLa talk.

This article will review three case studies of the use of PL theory for law, and then present ten promising avenues of potential research. Some of them are being actively pursued, others are promising speculations. I hope to make the case for the value of bringing programming-language methods to bear on legal problems, and to provide some notable examples of such problems as an enticement to researchers.

\section{Case Studies}

To understand what PL theory and law could do, we should start with what they have already done. Here are three examples of notable applications of programming-languages methods to law.

\subsection{Legal Logic Programming Languages}

There is an extensive subfield of artificial intelligence and law.\cite{mccarty1976reflections,welch1981lawgical,finan1981lawgical,sergot1986british,bench1987logic,gruner1989sentencing,debessonet1986artificial,dung2011modular,rissland1990artificial,prakkensartor2015,ashley2017artificial,livermore2019law} Its research program goes back nearly half a century, and it has its own dedicated association (the International Association for Artificial Intelligence and Law), conference series (the International Conference on Artificial Intelligence and Law), and journal (\emph{Artificial Intelligence and Law}). One of the major lines of research in AI and law is \emph{legal logic programing}: representing formal deductive legal reasoning using propositions expressed in a logic-programming language.\cite{sergot1986british, borrelli1989prolog,
holzenberger2020dataset,mccarty1976reflections}

As researchers have discovered, however, general-purpose logic programming
languages do not always capture the distinctive characteristics of legal
reasoning. For example, many legal conclusions are \emph{defeasible}: they are
valid on the basis of present knowledge, but could fail if a known exception
turns out to be the case.\cite{lawsky2017nonmonotonic} But classical logic is
monotonic: once a conclusion is established, it can never be falsified by
additional hypotheses. For this reason, it can be hard to model legal reasoning
in languages like Prolog whose semantics are monotonic. Capturing defeasible
reasoning (for example in statute definitions with
exceptions\cite{lawsky2017logic}) can require cumbersome circumlocutions. 

Thus,
some researchers have developed their own legal  programming languages by adding
defeasibility as a core language feature. Catala provides  a clean and rigorous
semantics for defeasible reasoning, which it uses to model statutory
definitions with exceptions.\cite{merigoux2020catala} PROLEG implements default
reasoning (in which conclusions have a default value even in the absence of
evidence), which it uses to reason about issues governed by burdens of
proof.\cite{satoh2010proleg} 

Other legal logic programming languages add different features. s(LAW) models legal ambiguity and administrative discretion by generating multiple answers, depending on the resolution of relevant ambiguities.\cite{arias2021modeling} LLD (a "Language for Legal Discourse") is a generic logic programming language enriched with temporal and deontic operators for reasoning about actions and obligations.\cite{mccarty1989language, mccarty2021position}

One could regard these examples merely as attempts to solve problems in AI and law. But I think it is telling that in so many cases the best way to solve a problem in AI and law has been to \emph{create a programming language}. It is a sign that there are aspects of legal reasoning that programming languages are uniquely well-suited to capture.

\subsection{The French Tax Code}
\label{sec:frenchtax}

The French tax code consists of about 3,500 pages of text, which the French tax authority (the DGFiP) must translate into an amount due for each taxpayer. Like any other large modern bureaucracy with millions of tasks, it does so by means of a computer program. But this program "relies on a legacy custom language [M] and compiler originally designed in 1990, which unlike French wine, did not age well with time."\cite{merigoux2020modern} The system is fragile and hard to maintain, and may contain bugs and mistakes.

A group of researchers are working with the DGFiP to transition its tax-computation system to a modern programming language backed up by a modern toolchain. They have reverse engineered M and given it a formal semantics, developed a new language (M++) with better formal properties and cleaner syntax, written a compiler (MLang) that supports both M and M++, and are using it to speed up, audit, debug, and ultimately \emph{prove correct} the algorithmic implementation of the tax laws.

This is a classic programming-languages story about replacing an \emph{ad hoc} system with one that rests on a firm theoretical and engineering foundation. M++ has rigorously specified semantics and its entire toolchain is built to support these semantics.

There are two levels here. The surface level is that MLang will make tax computations more reliable. But the deeper level is that because M++ makes it possible to express French tax law in a new way, it also makes it possible to \emph{understand}, \emph{reason about}, and \emph{debate} French tax law in new ways. Like law and economics, law and PL theory introduces new modeling tools that provide fresh perspectives on legal issues.

\subsection{Future Interests}

Orlando is a programming language for property conveyances that create future interests, with an accompanying implementation called Littleton.\cite{onward,grimmelmann2022language} Orlando's surface syntax is natural-language-like: it consists of styled conveyances like \texttt{O conveys to A for life, then to B so long as B does not marry}. Littleton then compiles these "programs" to a tree-based data structure modeling the different interests in a piece of property (here there are three A's life estate, B's remainder, and O's possibility of reverter). Orlando has an operational semantics that specifies how those interests change over time in response to events (e.g. \texttt{A dies}). And Littleton can reason about the interests in an Orlando tree to do things like correctly name interests and apply the Rule Against Perpetuities. Littleton is is not the only tool or heuristic for visualizing future interests,\cite{bayern2010formal, andersen1995present, reutlinger1994words} but it uses the formal power of a programming-language approach to provide a clean and extensible system.
 
On the most basic level, Orlando is useful as a teaching tool for exploring the consequences of particular conveyances. A student can tweak a conveyance to see what difference a wording change makes, or try out different sequences of events to see how the state of title evolves, or even turn on and off doctrines like the Rule in Shelley's Case. A teacher can put examples up on the screen and walk through them step by step, instantly updating based on questions and hypotheticals.

But on a deeper level, Orlando's syntax and semantics \emph{themselves} are are a scholarly claim about the underlying structure of the law of future interests. They are based on the rules codified in the Restatement (First) of Property\cite{rest-prop1} and glossed in generations of treatises and study aids.\cite{powell1949,berginhaskell,edwards, wendel2007treatise, sprankling2012understanding} They make precise, testable claims about property doctrines. Disagreements about property law can be reformulated as different rules in a formal semantics.


\section{Research Topics}

Let us turn, then to what PL theory has to offer to law -- and what law has to offer to PL theory. These research topics are arranged roughly in order from the most specific, concrete, and computational to the most general, abstract, and theoretical.

\subsection{Legal Domain-Specific Languages}

\textbf{New programming languages can provide formal models of specific legal subfields.}

 While the most familiar programming languages, like Java and C, are
 general-purpose langauges that can be used for a wide range of problems, a
 \emph{domain-specific language} (DSL) is specialized to meet the challenges of
 a specific problem domain.\cite{dslbib,fowler2010domain} For example, the R
 programming language for statistical analysis and data visualization includes
 operators for performing calculations on arrays of data, while the Ink
 programming language for interactive fiction includes features that select
 which scene the player will experience next.

"Law" as a whole is much too large for a DSL (at least for now), but many legal subfields are the right size for a DSL. PROLEG, LLD, and s(LAW) are programming languages for legal logic programming. M++ is a DSL for tax law. Orlando is a DSL for future interests in property law. 

The areas of law most amenable to formalization with DSLs are those where legal texts already already partly program-like. They are distinguished by three features. First they use \emph{rules rather than standards}, so that they are amenable to formalization at all. Second, their participants \emph{highly value clarity}, so that formalization offers real benefits. And third, they have \emph{recurring patterned structures}, so that programming-language tools capture their modular generativity.\cite{smith2006modularity, grimmelmann2022language}

Transactional private-law fields are the most obvious low-hanging fruit. It is not a coincidence that there are now numerous contract DSLs.\cite{jones2000composing,azzopardi2016contract,bahr2015,andersen2006compositional,findsl} Will drafting, for example, often involves the creation of trusts with common structures that have to be plugged together to fit the details of different family situations. Similarly, IP licensing breaks up rights into fairly standard elements, but the precise combination varies extensively from deal to deal. And many parts of property law have an underlying doctrinal structure that is itself recurring and patterned; programming languages for subareas like title assurance and land-use planning would take advantage of that structure.

\subsection{Hybrid Contracts}

\textbf{Programming languages for contracts can combine the advantages of legally enforceable human-readable contracts and technically enforceable machine-readable code.}
 
One of the most promising domains for legal DSLs is contracts, in part because here is where the most work has already been done. There has been extensive research products to model the deontic, multi-party, and event-driven structure of contracts.\cite{azzopardi2016contract, azzopardi2018observing, camilleri, bahr2015, andersen2006compositional} More recently, the rise of blockchain-based smart contracts has driven extensive interest in developing good PLs for expressing them.\cite{ladleif2019unifying, solidity,accord}

This last convergence suggests that there may be substantial conceptual and practical payoffs to developing systems that seamlessly hybridize human-readable terms and machine-readable logic. Users of such a system could write contractual terms once and them compile them both to "legal code" for humans and "computer code" for automated execution.\cite{surden2012computable} This approach could help solve one of the most vexing problems in smart-contract drafting: what to do when the actual code of a smart contract and its human-readable summary diverge?\cite{levine2016blockchain}

Another advantage of hybrid contracts is that formalizing the formalizable parts of them would help prevent -- or at least detect -- bugs in their logic. A human might miss an inconsistency in the natural-language version of a payment schedule, but a computer running the right algorithm over a software-language version of the same schedule would have a chance at flagging the problem. In addition, it would become more possible to automate execution of \emph{part} of a contract, with responsibility for its successful performance being handled collaboratively by computers and humans.

More broadly, having a good programming-language-driven approach to hybrid contracts would help enormously in solving the difficult doctrinal problems they create.\cite{allen2018wrapped, cohneytransactional, grimmelmann2019all} It would help lawyers and judges conceptualize and answer questions about the legal effects of executing the automated part, and about the technical effects of legal enforcement.

\subsection{Orthogonal Legal Primitives}  

\textbf{Some legal concepts are so simple and so pervasive that they can serve as building blocks for legal programming languages.}

There is a striking underlying simplicity in attempts to model distinctive features of legal reasoning. Orlando compiles an enormous range of property conveyances like "to Alice for life, then to Bob" to a small number of operators. It boils down the \emph{effects} of a conveyance (rather than the \emph{language} of the conveyance itself) into (1) the \emph{termination} of interests under specified conditions ("to Alice \emph{for life}"), and (2) the \emph{sequencing} of interests ("to Alice \ldots \emph{then} to Bob"). Similarly, although Catala has a rich syntax of scopes, contexts, and definitions to allow users to track the structure of statutes, its semantics are built around extending standard lambda calculus with a single new feature: exceptions. It is defaults done right for the lambda calculus.

Beneath each of these legal programming languages, then, is a legal calculus: "a language with well-defined syntax and semantics that has features whose main purpose is to model particular aspects of its [legal] target domain."\cite{basu2022calculi} The crown jewel of legal calculi is the Haskell-based  language used by Jones, Eber, and Seward to model option contracts in their article \emph{Composing Contracts}, which has combinators for fundamental contractual operations like allowing a party to choose one of two obligations to take on (\texttt{or}) and allowing a party to choose when to take on an obligation (\texttt{anytime}).\cite{jones2000composing}
 Its minimal set of ten combinators is both powerful and elegant; they carve up option contracts along the joints.

There is something fruitful about the exercise of identifying a legal concept and isolating it as a specific language feature. In the programming-language world, a language's elegance is often thought of in terms of the degree to which it is built from  \emph{orthogonal primitives}.\cite{landin1966next,steele1998growing} Each of these features should be "primitive" in that it cannot be further decomposed, and the features should be "orthogonal" to each other in that they have simple and easily predictable interactions.

A profitable line of research would be to identify additional legal primitives and to systematize their relationships to each other. Existing examples already discussed include the sequencing and termination of legal interests (as in Orlando), the conjunction and disjunction of legal relationships (as in \emph{Composing Contracts}), defaults and exceptions (as in Catala and PROLEG), and legal obligations and entitlements (as in LLD). This last example has a particularly distinguished heritage in law, as Hohfeld's system of jural opposites and jural correlatives is in a sense an informal and very early legal calculus.\cite{hohfeld1913applied, hohfeld1917fundamental} It has been the foundation for numerous efforts in AI and law,\cite{allen1974formalizing} and the interlocking nature of claim-rights and duties is essential to a programming language for contracts of any complexity. Other legal concepts that might usefully be translated into programming-language features include the creation and combination of entities in business associations law, hierarchies of authority among sources of law, the movement of jurisdiction over a case among courts during litigation, counterfactuals and hypotheticals, and concepts of transfer and notice in commercial law.

Any such effort needs to answer similar questions. First and foremost is the quest for clean minimal formalizations. This can be a difficult enterprise.  Jones, Eber, and Seward observe:
\begin{quote}
Identifying the "right" primitive combinators is quite a challenge. For example, it was a breakthrough to identify and separate the two forms of choice {\texttt{or}} and {\texttt{anytime}}, and encapsulate those choices (and nothing else) in two combinators.
\end{quote}
Second, there is orthogonality. Consider, for example, the problem of combining deontic operators with operators for transferring property. Do the transferor's obligations "run with the land" to a transferee? In law, the answer is sometimes yes and sometimes no. Thus, obligation and transfer are not fully orthogonal; any system with both sets of operators must also have mechanisms to deal with their interaction by marking which obligations do and do not follow an interest in property. This is a hard problem, and that is part of what makes it interesting.

Third, the possibility that many legal concepts can be realized as programming-language primitives implies that \emph{different} legal PLs will often have \emph{common} features. Thus, perhaps it would be useful to build legal-specific back-end support into the programming-language toolchain, not to support any particular legal programming language but instead to support a range of projects using different legal PLs. At the moment, the most common way to implement a legal programming language is to compile it to an off-the-shelf back-end. Is there enough in common among different legal PLs to justify adding law-specific features (e.g. for defaults or obligations) to one of those off-the-shelf back ends? I do not know the answer, but the question seems worth asking.

\subsection{Legal Drafting Languages}

\textbf{Legal drafting could be made clearer and less error-prone by incorporating concepts from PL theory.}

Programming-language theory helps programmers write better programs by giving them languages with features that promote clear and correct code. Perhaps programming-language theory can do the same for law. While some legal PLs are designed to model legal relationships that already exist, programming-language concepts could also be useful in improving legal drafting, where clarity and correctness are already important goals.\cite{stark2013drafting}

For example, Lawsky observes that the scope of statutory definitions is often ambiguous.\cite{lawsky2016formalizing} To be sure, some legal language is deliberately vague (e.g., "within a reasonable time"), but the scope of a statutory definition should be wholly unambiguous: either the definition of "home equity indebtedness" under the Internal Revenue Code should include the  1 million limitation on "acquisition indebtedness" or it should not. But the way the definitions are drafted does not resolve the issue. Lawsky proposes that legislative drafters should formalize their statutory definitions using first-order predicate logic, and then use those formalizations to guide their legislative drafting to make explicit the intended scope of each definition.

In other words, Lawsky proposes using first-order logic as a kind of \emph{programming language for legal drafting}. This language has a feature -- explicit scoping -- that promotes clear and correct legal code. It is actually impossible to draft a tax provision in first-order logic with ambiguous scope, because that provision will violate simple and easily checked syntactic requirements that the language imposes on programs written in it. And once the provision is drafted in a logical form, it can then be "compiled" to the native format of the legal system -- natural language -- through a straightforward process that hopefully will not introduce errors of this sort. While, other scholars have addressed the value of logical formalization for legal reasoning and legal drafting, \cite{allen1977normalized} Lawsky's analysis focuses attention on the way in which this formalization benefits from an appropriate programming language.

The power of this example prompts the question: \emph{what other programming-language features would promote clear and unambiguous drafting}? A few possible answers include:

\begin{description}
\item[Variables and binding] The inverse problem of specifying what scope of use a given definition applies to is specifying what binding of a variable a given reference refers to. PL theory has a rich collection of concepts (e.g., lexical versus dynamic scope), algorithms (e.g., data-flow analysis), and programming techniques (e.g., marking variables as global) to manage this problem. Legal drafters have their own rules of thumb to manage the problem. For example, patent drafters write "a" when introducing a new claim element and to write "the" when referring to that element subsequently, to make clear the distinction between binding and reference.
\item[Cross-referencing] Statutes and contracts are liberally sprinkled with cross-references (e.g. "for the purposes of subparagraph (e), 'days' shall mean business days"). But cross-references are fragile as texts are amended; drafters frequently fail to catch all of the references to a renumbered section, resulting in mistaken or nonsensical cross-references. This too is a problem that programmers contend with, and numerous programming-language features promote correct coordination between different parts of a program. Function declarations, module systems, preprocessor includes, and type checking all enable coordination while trying to prevent the kinds of mistakes that it can introduce.
\item[Counterfactuals and hypotheticals] Legal drafting is full of counterfactuals and hypotheticals. For example, a party may be obligated to take action $x$ if failing to do so \emph{would result in} some state of affairs $y$. This type for hypothetical reasoning is one of the basic skills taught in law school, and lawyers rely on it extensively. programming-language theory has tools for reasoning about possible executions of a program, such model checking. And some language features, such as reflection, enable programs to reason about themselves. A good legal drafting programming language might include features for compactly stating counterfactuals and hypotheticals.
\item[Substitution] Lawyers do search-and-replace all the time when they are drafting or amending legal documents. Sometimes this is explicitly a feature of the documents themselves, as when "in the case of an Adverse Event, all deadlines hereunder shall be extended for ten days." Substitution is absolutely fundamental to lambda-calculus based semantics, and PL theory has both a rich mathematical theory of substitution and an extensive arsenal of techniques and tools to employ substitution in a principled way when constructing and reasoning about programs.
\item[Abstraction] From polymorphism to virtual base classes, from abstract data types to functors, PL theory as a field is characterized by the deployment of powerful abstractions that massively generalize common programming patterns. Legal drafting's embrace of abstraction has been much more tentative; lawyers find themselves writing the same legal code over and over with minute variations. Good legal programming-language techniques for managing systematic abstractions could help lawyers avoid the tedium and risk of error that come from doing a hundred times what could be done once.
\end{description}


\subsection{Legal Design Patterns}

\textbf{Common patterns in legal drafting and design should be systematically isolated and described, much as common patterns in software design already are.}

In software engineering, a \emph{design pattern} is a general, reusable template for solutions to a commonly occurring class of design problems.\cite{gamma1995elements} The problems are not standardized enough that one can use a literally identical solution: if that were the case, one could simply reuse an existing library. Nor are they so diverse that each requires a bespoke solution from scratch. Rather, design patterns are useful in the middle ground, where the same kinds of problems recur, but the details are a little different each time. If this sounds a lot like legal work, you are not wrong.

The roots of design patterns lie in the work of Christopher Alexander and his collaborators in architectural theory; their 1977 book \emph{A Pattern Language} described patterns like "light on two sides of every room" and argued that they fit into an interlocking, mutually reinforcing structure.\cite{alexander1977pattern} These ideas were carried over into the software engineering literature by developers and designers who appreciated its connections to modularity and the way in which object-oriented programming, in particular, could be used to implement patterns.\cite{gabriel1996patterns, gamma1995elements} There and in related fields, such as user interface design, they have built up rich libraries of design patterns.\cite{, crumlish2009designing, tidwell2010designing} There is a close connection between design patterns and PL theory, because languages are frequently created to make certain types of design patterns easy to employ.

There is a nascent literature on legal design patterns, which identifies some common patterns in legal drafting.\cite{gerding2013contract, roach2016toward, cunningham2006language} This literature draws both on the original architectural theory and on the idea of design patterns in software.

A more ambitious agenda would seek to craft pattern languages for specific legal areas. In a sense, this is what some of the the formbooks and drafting manuals used by legal practitioners are getting at: they break contracts, licenses, settlements, security interests, regulations, and many other types of legal documents down into semi-standard parts, explain what each of those parts can do, and show how to fit them together. But this process can be systematized and scaled up. What are the proxies, decorators, and singletons in complex corporate transactions? What is the complete pattern language of will drafting?

\subsection{Design Principles}

\textbf{Broad design principles from programming-language theory are recognizable throughout the law.}

Design patterns are solutions to specific commonly occurring problems. At a higher level of abstraction are design \emph{principles}: general features of good software design that promote reliable, maintainable, and efficient code. 

A key example is \emph{modularity}: the decomposition of a system into subsystems that are weakly coupled to each other. The theory of modularity is rooted in cybernetics,\cite{simon1996sciences} cite but has become a central principle of computer system design.\cite{baldwin2000design}  It is often achieved with explicit programming-language features. Programming-language support for separate compilation modules, for example, is literally about modularity: it's right there in the name. Other examples include functions,  abstract data types, and objects.

Modularity in particular has already been fruitfully applied to legal theory. Henry Smith, who holds a Ph.D. in linguistics and draws heavily on the cybernetic theory of modularity, has shown how the boundaries of property (both physical and intellectual) are often drawn in modular ways to manage the information costs of dealing with other people's rights in things. \cite{smith2011standardization, optimalstandardization, lawofthings} He and other scholars have applied modularity theory to legislative drafting, contracts, and internet regulation. \cite{hwang2015unbundled,hwang2018deal,yoo2016modularity,van2012internet, blackwell1999finally, smith2006modularity}

Other examples of broad programming-language design principles that may be relevant for law include recursion, compsitionality, type-safety, and extensibility. Legal drafting in particular already makes \emph{ad hoc} use of all of these principles, and systematizing them could help drafters be more precise. Ideal contract terms are not just modular but compositional: combining them produces no unexpected interactions, and replacing one adverse-events clause with another, say, should affect other parts of the contract. The breakdown of a contract into different kinds of clauses (representations versus promises, for example) is about typing, and a type-safe drafting framework would actually prevent drafters from using one in place of the other by mistake.

Corporate law uses modularity in concentrating legal relationships within discrete units; Coase's \emph{Nature of the Firm} is in essence an argument about the optimal size of modules. It also makes heavy use of extensibility in providing a set of default classes for corporate forms that can be extended with custom legal logic. Commercial law's treatment of transferrable obligations is modular and recursive: the passage of warranties along a chain of transfers is a recursive solution to the trust problem of dealing with strangers. The Federal Rules of Evidence's open-ended delegation of authority to the federal courts to make privilege rules is an extensibility framework. Other examples await those who are willing to go in search of them.

\subsection{An IDE for Lawyers}

\textbf{Lawyers could benefit from drafting tools as sophisticated as those that programmers enjoy.}

There was a time when programming a computer meant punching holes in paper cards, flipping switches for every bit, or even hand-wiring the program into physical connections between components. But one of the great accomplishments of computer science has been the conversion of programming from an arduous and punishing task into a process for leveraging human creativity. The modern toolchain supporting programmers -- including tools for code editing, compilation, version control, code analysis, debugging, testing, and analytics -- is powerful and extensive.

At the center of the toolchain are Integrated Development Environments (IDEs) like Visual Studio Code, xCode, Eclipse, and Sublime Text. Consider just a few ways an IDE supports effective programming. As the programmer types, their code is automatically color-coded to keep the program's structure clear. The IDE  suggests sensible auto-completions based not just on a static dictionary but on the rest of the program's code. The programmer can click on any function or variable to get more information about it (e.g. its type) and jump to other places it is used. With another click, they can reformat their code to make it easier to read. They can browse past versions of a code block to see how it has changed over time, and save changes to a shared repository. They can launch the program, set a breakpoint, and step through the code line by line to see how it behaves and where it goes wrong.

Now compare the legal drafting toolchain. By far the most common IDE for lawyers is Microsoft Word, and the most common version-control system is saving drafts as files with names like \texttt{Licensing Agreement 3.2 MH Final 2 USE THIS ONE}. If you are lucky (e.g. because you worked in a legislative drafting office), you might use a specialized XML-based tool. If you are very unlucky, you are still stuck with WordPerfect. The state of the art for most lawyers is track changes and a handful of Word plugins to assist with tasks like formatting citations. Other parts of the toolchain are evolving quickly, especially case and document management. But the IDE lags behind.

What might a true language-aware programming-language-based IDE for lawyers drafting structured documents (e.g., statutes, contracts, and wills) look like? 
\begin{itemize}
\item It would feature syntax highlighting to distinguish different kinds of text, and automatic formatting for the structural framework of documents.
\item  It would use linting, type-checking, and other static analyses to enforce important invariants (e.g. that every defined term actually have exactly one definition) and to find common errors. 
\item It would integrate directly to version control systems with easy-to-use features like branches, diffs, merges, and pull requests.
\item  It would integrate into project management and issue-tracking systems so that team members could mark to-dos and file bug reports against specific blocks of text.
\item It would also be integrated into knowledge-management systems used by law firms, agencies, and other institutions tracking immense quantities of documents.
\item  It would perform unit tests against any changes to ensure that they don't break existing features, and carry out fuzz testing with possible sequences of events to ensure robustness against unexpected contingencies. 
\item It would provide interactive simulation and debugging features like visualization and breakpoints so that drafters could test out what the instrument they were creating will actually do in various situations.
\end{itemize}

\subsection{Jupyter Notebooks for Law}

\textbf{Interactive and visual tools for experimenting with code could be as powerful in law as they are in programming.}

A great advantage of programming-language approaches is that they often improve understanding. Subsymbolic AI is notorious for having an explainability problem,\cite{selbst2018intuitive, kaminski2019right, strandburg2019rulemaking} but for the most part not so PL theory. A good compiler's report of a type conflict directs the programmer's attention to a comprehensible and fixable mistake in their assumptions. Again, it is PL theory's embrace of the \emph{structure} of programs that provides a foundation for making that structure comprehensible.

Within law and PL theory, a number of projects aim to present legal structures in an especially human-comprehensible way. Some of them are teaching tools for students and professors in a classroom setting, others are modeling tools for practitioners in a real-world setting, a few are both. As Lawsky observes, the modern tax form is in a sense a synthesis of formal and informal approaches, but it should be possible to do much better.\cite{lawsky2020form}

One way in which programming-language techniques can help is visualization: although lawyers are notoriously verbal thinkers, sometimes a good diagram is worth a thousand words. For example, Littleton displays future interests using a "railroad diagrams" library originally created to show how a programming language's syntax works; Bayern's future-interests tool uses branching arrows.\cite{grimmelmann2022language,bayern2010formal} The two kinds of visualization are complementary, both to each other and to the textual formulations used by lawyers and law students. The use of programming-language techniques provides a more principled foundation than the informal diagrams drawn by generations of Property teachers.\cite{andersen1995present}

Another great advantage of programming-language techniques is interactivity. Unlike a pre-canned set of examples, they can adapt on the fly to the user's queries, allowing in-depth exploration and the ability to test one's understanding with variations. Littleton and Bayern's interpreter both allow users to try a conveyance, examine the output, vary its details, try it again, and so on, creating a tight feedback loop.\cite{grimmelmann2022language,bayern2010formal} Lawsky Practice Problems uses programming-language techniques to \emph{generate} an unlimited number of different tax practice problems, with the names, numbers, and doctrinal classification varying each time.\cite{lawskypracticeproblems}

Another way of integrating programming languages and law to improve understanding is through literate legal drafting. Literate \emph{programming} is a style of programming in which a natural-language explanation of a program's functionality is interleaved with and generates the code implementing that functionality.\cite{knuth1984literate} This vision inspires several recent projects in PL theory and law. The Catala project is working on having lawyers and programmers collaborate in a kind of pair programming, with the legal specification and program logic tightly integrated.\cite{merigoux2020modern, garber2020pair, williams2003pair} Similarly, the Accord Project's Cicero templating system is designed to embed machine-executable logic inline in the language of contracts.\cite{accord}

An ambitious goal combining visualization, interactivity, and literate programming would be to create the legal equivalent of a Jupyter notebook.\cite{jupyter} It would freely intermingle natural-language text addressed to humans with executable segments addressed to computers. It would let users visualize legal structures using programming-language concepts like abstract syntax trees and control-flow graphs. It would tighten the feedback loop between writing legal text and seeing what it does to a matter of seconds or less.

\subsection{The Law of Software}

\textbf{Programming-language theory can answer doctrinal questions about software by describing what programs are and how they work.}

Programming is not, like medicine, a regulated profession. But programs are
subject to law in many ways. Here are just a few of the many doctrinal questions
whose answers depend in part on questions about which programming-language
theorists have relevant expertise. 

\begin{itemize}
  \item The patentability of an invention turns on whether it is an "abstract idea," and if so, whether it adds something "significantly more" to the idea.\cite{alice2014} The inherent abstraction of software means that this two-step inquiry is in play in any case claiming software functionality.
  \item Also in patent law, infringement occurs when a defendant supplies  "components" of an invention in the United States for "combination" abroad." The Supreme Court held that a CD-ROM of Microsoft Windows was not a "component" of a patented invention, because only the abstract software on the CD-ROM, not the CD-ROM itself, was installed on computers to make an infringing combination.\cite{microsoft2007}
  \item The Copyright Act protects computer programs as literary works. But not all programming languages are straightforwardly textual. For example, in visual programming languages such as Scratch, a programmer manipulates graphical elements in two dimensions. And even in more traditional programming environments like Apple's Xcode, developers frequently combine program text with visual interface designs.
  \item Copyright protection does not extend to elements of a program that are "dictated by efficiency," by "external factors" such as compatibility with particular hardware, or by "widely accepted programming practices."\cite{altai1992}
  \item The Supreme Court has held that it is fair use to copy an application programming interface (API).\cite{oracle2021} Dividing a program into "interface" and "implementation" requires detailed engagement with the details of the language in which it is written, and determining what elements are required for compatibility also requires considering how that language is compiled or interpreted.
  \item The First Amendment covers the publication of software as a means of communicating ideas to other programmers.\cite{junger2000} But using software for its functional effects, for example to spy on one's spouse's computer,  is not automatically protected. Drawing the line requires a theory of what constitutes the expressive speech in software.\cite{tien2000publishing}
  \item Similarly, there may be a right \emph{not} to write software; Apple has argued that being required to write code to unlock an iPhone would violate the First Amendment right against compelled speech. This argument would obviously fail if Apple were required to open a physical vault, so again one needs a theory of what is expressive and what is functional about software.
  \item Smart contracts are intended to supplement or replace legally binding contracts. Courts have a rich theory of how to interpret human-readable legal contracts, and they will need a similarly rich theory of how to interpret machine-readable smart contracts.\cite{allen2018wrapped, klass2022vending,grimmelmann2019all}
  \item The Computer Fraud and Abuse Act prohibits some uses of computers "without authorization." Under some circumstances, what counts as "authorization" is determined by the program itself.\cite{grimmelmann2013casino,grimmelmann2016consenting} (For example, a website authorizes or prohibits access by comparing a user-entered password to the password stored in its user database.) Again, there are interpretive questions that one needs a theory of computer-program meaning to answer.
\end{itemize}

\subsection{Philosophical Questions}

\textbf{Lawyers and programmers can learn from each other what it means to interpret a text.}

Legal and programming-language scholars both study interpretation: how to understand what a text means. The kinds of texts they study are different, but as we have seen repeatedly, this shared focus on text and its meaning is a large part of what makes their collaboration so fruitful. Indeed, this shared focus on text is \emph{itself} a topic of interest. A deeper understanding of interpretation is not just of practical use in applying programming-language methods to legal problems: it can help theorists in both fields understand their fields better.

As we have seen, many doctrinal questions in PL theory and law are at the boundary between computer science and philosophy: what is a program, how does it work, and what does it mean? For example, the claim that software consist entirely of mathematics and is therefore not properly patentable or copyrightable\cite{klemens2005math,moglen1999anarchism} is grounded in programming-language theories of the semantics of programming languages. But this is not the only possible way to conceive of software,\cite{mackenzie2001mechanizing} and the field of PL theory and law has much to contribute in complicating, challenging, and fleshing out this picture. 

The fundamental theoretical question of PL theory and law is \emph{what is the difference between how people and computers interpret texts}? Answers to this question tell us both what law and programming languages can do (because there are relevant similarities) and what it cannot (because there are unbridgeable differences). It tells us about what is truly fundamental about legal interpretation, and what is a historical accident of the fact that it developed in a pre-computer age. It tells us what changes about a rule when it is formalized, and it tells us about what the Church-Turing thesis means for society.

The fundamental practical question of PL theory and law is \emph{how (if at all) should the availability of computers change how legal texts are written and interpreted}? I have argued that some (not all) legal rules should be made more formal, and that the legal drafting process should be more like the software development process.  One could go much further: delegating more of the drafting process to computers, and using algorithms to interpret and apply laws.\cite{solum2014artificial, casey2016self, casey2017self, ben2021personalized} Of course, just because something can be done, doesn't mean it should.\cite{burk2019algorithmic, klass2022tailoring} But this is a question that the legal profession urgently needs to ask itself, and programming-language scholars and software developers are in a prime position to help think through the possibilities and their consequences.

\section{Closing Thoughts}

Building a field of law and programming languages will not be easy; interdisciplinary work never is. Beyond the obvious point that each field has its own rich set of concepts and immense literature, the disciplines differ in more subtle ways. They have different research methods: 

They have different research methods: programming-language researchers mostly build new things in teams, while legal researchers mostly study existing things alone. They have different standards of rigor: to caricature just a little, a programming-languages paper is done when the proofs are filled in, a law paper when the footnotes are filled in. They have different authorities: "because Congress said so" is a good answer in law, but not in programming languages. And they have different writing styles: to a programming-language scholar, the typical law-review article is shockingly long and repeats far too much that everyone already knows, while to a legal scholar, the typical programming-language paper is shockingly short and omits essential background and context.

One particular challenge is that scholars can be overconfident amateurs outside of their own field. The phrase "law-office history" is usually a pejorative: it implies that legal scholars do poor history because they lack the training, the patiences, and the motivation to get history right on its own terms. Of course, the reverse is also true: microeconomists sometimes get a bad rap in the legal academy for bursting in to long-running debates like the Kool-Aid Man, supremely confident that their stylized models hold all the answers.

Some of the answers to these challenges are the same in programming languages and law as they are in any interdisciplinary project. Collaboration will be essential, and some of the most interesting research efforts in the space involve teams with lead researchers from both programming languages and law. In addition, dual training in law and CS, although rare, is immensely helpful in bridging the two fields. There is a slow but growing trickle of researchers with knowledge of both, who have an important role in training the next generation of law and programming-language researchers.\cite{shadab2020software, ohm2009computer}

I wish to close where I started, by asking again the question, why these two fields? Why law and programming languages? What do they have in common that other pairs of fields do not?

The answer, as I have hinted at throughout, is that law and programming languages share a common focus on how professionals use precisely structured linguistic constructions to do things in the world. Law has good reason to be interested in many CS fields, from cryptography to machine learning, but within CS, it is programming languages that most shares law's linguistic focus. And programming-language techniques can usefully be applied to many problem domains, but law happens to be particularly rich in the kind of problems for which programming-linguistic solutions can be useful. Historians and sociologists have their own interesting and important problems, but by and large they are not the kind of problems that a programming language can help solve.

If there is a lesson to take from the history of law and programming languages's engagement, it is that principled methods are often superior to \emph{ad hoc} ones. programming languages has a long and impressive history of developing concepts, languages, and tools to tame the chaos of coding. Law and legal tech, which are engaged in their own eternal struggle to clean up the Augean stables of law and life, should welcome all the help they can get. Formalizing a body of law well enough to turn it into a programming language forces you to understand it in a far deeper way -- and it is that hard-earned insight that law and programming languages promises.

\begin{acks}
This work was supported by NSF Award FMitF-2019313. My thanks to the organizers of and participants in the ProLaLa 2022 workshop, where I delivered an earlier version of this article as a talk, and to Aislinn Black and Sarah Lawsky.
\end{acks}

\bibliographystyle{ACM-Reference-Format}
\bibliography{refs}


\begin{thebibliography}{114}


\ifx \showCODEN    \undefined \def \showCODEN     #1{\unskip}     \fi
\ifx \showDOI      \undefined \def \showDOI       #1{#1}\fi
\ifx \showISBNx    \undefined \def \showISBNx     #1{\unskip}     \fi
\ifx \showISBNxiii \undefined \def \showISBNxiii  #1{\unskip}     \fi
\ifx \showISSN     \undefined \def \showISSN      #1{\unskip}     \fi
\ifx \showLCCN     \undefined \def \showLCCN      #1{\unskip}     \fi
\ifx \shownote     \undefined \def \shownote      #1{#1}          \fi
\ifx \showarticletitle \undefined \def \showarticletitle #1{#1}   \fi
\ifx \showURL      \undefined \def \showURL       {\relax}        \fi
\providecommand\bibfield[2]{#2}
\providecommand\bibinfo[2]{#2}
\providecommand\natexlab[1]{#1}
\providecommand\showeprint[2][]{arXiv:#2}

\bibitem[Accord Project(2022)]%
        {accord}
Accord Project \bibinfo{year}{[2022]}\natexlab{}.
\newblock \bibinfo{booktitle}{\emph{{Accord Project}}}.
\newblock Accord Project.
\newblock
\urldef\tempurl%
\url{https://docs.accordproject.org}
\showURL{%
\tempurl}


\bibitem[Alexander et~al\mbox{.}(1977)]%
        {alexander1977pattern}
\bibfield{author}{\bibinfo{person}{Christopher Alexander},
  \bibinfo{person}{Sara Ishikawa}, \bibinfo{person}{Murray Silverstein},
  \bibinfo{person}{Max Jacobson}, \bibinfo{person}{Ingrid Fiksdahl-King},
  \bibinfo{person}{Angel Shlomo}, {et~al\mbox{.}}}
  \bibinfo{year}{1977}\natexlab{}.
\newblock \bibinfo{booktitle}{\emph{A Pattern Language: Towns, Buildings,
  Construction}}. Vol.~\bibinfo{volume}{2}.
\newblock \bibinfo{publisher}{Oxford University Press}.
\newblock


\bibitem[Allen(2018)]%
        {allen2018wrapped}
\bibfield{author}{\bibinfo{person}{Jason~G. Allen}.}
  \bibinfo{year}{2018}\natexlab{}.
\newblock \showarticletitle{Wrapped and Stacked: ‘Smart Contracts’ and the
  Interaction of Natural and Formal Language}.
\newblock \bibinfo{journal}{\emph{Eur. Rev. Cont. L.}} \bibinfo{volume}{14},
  \bibinfo{number}{4} (\bibinfo{year}{2018}), \bibinfo{pages}{307}.
\newblock


\bibitem[Allen(1974)]%
        {allen1974formalizing}
\bibfield{author}{\bibinfo{person}{Layman~E. Allen}.}
  \bibinfo{year}{1974}\natexlab{}.
\newblock \showarticletitle{Formalizing Hohfeldian Analysis to Clarify the
  Multiple Senses to Legal Right: A Powerful Lens for the Electronic Age}.
\newblock \bibinfo{journal}{\emph{S. Cal. L. Rev.}}  \bibinfo{volume}{48}
  (\bibinfo{year}{1974}), \bibinfo{pages}{428}.
\newblock


\bibitem[Allen and Engholm(1977)]%
        {allen1977normalized}
\bibfield{author}{\bibinfo{person}{Layman~E. Allen} {and}
  \bibinfo{person}{C.~Rudy Engholm}.} \bibinfo{year}{1977}\natexlab{}.
\newblock \showarticletitle{Normalized Legal Drafting and the Query Method}.
\newblock \bibinfo{journal}{\emph{J. Legal Educ.}}  \bibinfo{volume}{29}
  (\bibinfo{year}{1977}), \bibinfo{pages}{380}.
\newblock


\bibitem[Andersen et~al\mbox{.}(2006)]%
        {andersen2006compositional}
\bibfield{author}{\bibinfo{person}{Jesper Andersen}, \bibinfo{person}{Ebbe
  Elsborg}, \bibinfo{person}{Fritz Henglein}, \bibinfo{person}{Jakob Grue},
  {and} \bibinfo{person}{Christian Stefansen}.}
  \bibinfo{year}{2006}\natexlab{}.
\newblock \showarticletitle{Compositional Specification of Commercial
  Contracts}.
\newblock \bibinfo{journal}{\emph{Int'l J. on Software Tools for Tech.
  Transfer}}  \bibinfo{volume}{8} (\bibinfo{year}{2006}), \bibinfo{pages}{485}.
\newblock


\bibitem[Andersen(1995)]%
        {andersen1995present}
\bibfield{author}{\bibinfo{person}{Roger~W. Andersen}.}
  \bibinfo{year}{1995}\natexlab{}.
\newblock \showarticletitle{Present and Future Interests: A Graphic
  Explanation}.
\newblock \bibinfo{journal}{\emph{Seattle U. L. Rev.}}  \bibinfo{volume}{19}
  (\bibinfo{year}{1995}), \bibinfo{pages}{101}.
\newblock


\bibitem[Arias et~al\mbox{.}(2021)]%
        {arias2021modeling}
\bibfield{author}{\bibinfo{person}{Joaqu{\'\i}n Arias}, \bibinfo{person}{Mar
  Moreno-Rebato}, \bibinfo{person}{Jose~A Rodriguez-Garc{\'\i}a}, {and}
  \bibinfo{person}{Sascha Ossowski}.} \bibinfo{year}{2021}\natexlab{}.
\newblock \showarticletitle{Modeling administrative discretion using
  goal-directed answer set programming}. In
  \bibinfo{booktitle}{\emph{Conference of the Spanish Association for
  Artificial Intelligence}}. \bibinfo{publisher}{Springer},
  \bibinfo{pages}{258--267}.
\newblock


\bibitem[Ashley(2017)]%
        {ashley2017artificial}
\bibfield{author}{\bibinfo{person}{Kevin~D. Ashley}.}
  \bibinfo{year}{2017}\natexlab{}.
\newblock \bibinfo{booktitle}{\emph{Artificial Intelligence and Legal
  Analytics: New Tools for Law Practice in the Digital Age}}.
\newblock \bibinfo{publisher}{Cambridge University Press}.
\newblock


\bibitem[Azzopardi et~al\mbox{.}(2018)]%
        {azzopardi2018observing}
\bibfield{author}{\bibinfo{person}{Shaun Azzopardi}, \bibinfo{person}{Gordon~J.
  Pace}, {and} \bibinfo{person}{Fernando Schapachnik}.}
  \bibinfo{year}{2018}\natexlab{}.
\newblock \showarticletitle{On Observing Contracts: Deontic Contracts Meet
  Smart Contracts}. In \bibinfo{booktitle}{\emph{Proc. 31st Int'l Conf. on
  Legal Knowledge \& Info. Systems (JURIX 2018)}}. \bibinfo{publisher}{IOS
  Press}, \bibinfo{pages}{21}.
\newblock


\bibitem[Azzopardi et~al\mbox{.}(2016)]%
        {azzopardi2016contract}
\bibfield{author}{\bibinfo{person}{Shaun Azzopardi}, \bibinfo{person}{Gordon~J.
  Pace}, \bibinfo{person}{Fernando Schapachnik}, {and} \bibinfo{person}{Gerardo
  Schneider}.} \bibinfo{year}{2016}\natexlab{}.
\newblock \showarticletitle{Contract Automata}.
\newblock \bibinfo{journal}{\emph{Artificial Intelligence \& L.}}
  \bibinfo{volume}{24} (\bibinfo{year}{2016}), \bibinfo{pages}{203}.
\newblock


\bibitem[Bahr et~al\mbox{.}(2015)]%
        {bahr2015}
\bibfield{author}{\bibinfo{person}{Patrick Bahr}, \bibinfo{person}{Jost
  Berthold}, {and} \bibinfo{person}{Martin Elsman}.}
  \bibinfo{year}{2015}\natexlab{}.
\newblock \showarticletitle{Certified Symbolic Management of Financial
  Multi-Party Contracts}. In \bibinfo{booktitle}{\emph{Proc. 20th ACM SIGPLAN
  Int'l Conf. on Functional Programming}}. \bibinfo{publisher}{Association for
  Computing Machinery}, \bibinfo{pages}{315}.
\newblock


\bibitem[Baldwin and Clark(2000)]%
        {baldwin2000design}
\bibfield{author}{\bibinfo{person}{Carliss~Y. Baldwin} {and}
  \bibinfo{person}{Kim~B. Clark}.} \bibinfo{year}{2000}\natexlab{}.
\newblock \bibinfo{booktitle}{\emph{Design Rules: The Pwer of Modularity}}.
\newblock \bibinfo{publisher}{MIT Press}.
\newblock


\bibitem[Basu et~al\mbox{.}(2019)]%
        {onward}
\bibfield{author}{\bibinfo{person}{Shrutarshi Basu}, \bibinfo{person}{Nate
  Foster}, {and} \bibinfo{person}{James Grimmelmann}.}
  \bibinfo{year}{2019}\natexlab{}.
\newblock \showarticletitle{Property Conveyances as a Programming Language}. In
  \bibinfo{booktitle}{\emph{Proceedings of the 2019 ACM SIGPLAN International
  Symp. on New Ideas New Paradigms \& Reflections on Programming \& Software
  (Onward!)}}. \bibinfo{publisher}{Association for Computing Machinery},
  \bibinfo{pages}{128}.
\newblock


\bibitem[Basu et~al\mbox{.}(2022)]%
        {basu2022calculi}
\bibfield{author}{\bibinfo{person}{Shrutarshi Basu}, \bibinfo{person}{Anshuman
  Mohan}, \bibinfo{person}{Nate Foster}, {and} \bibinfo{person}{James
  Grimmelmann}.} \bibinfo{year}{2022}\natexlab{}.
\newblock \showarticletitle{Legal Calculi}. In
  \bibinfo{booktitle}{\emph{Programming Languages and the Law (ProLaLa)}}.
  \bibinfo{publisher}{Association for Computing Machinery}.
\newblock


\bibitem[Bayern(2010)]%
        {bayern2010formal}
\bibfield{author}{\bibinfo{person}{Shawn~J. Bayern}.}
  \bibinfo{year}{2010}\natexlab{}.
\newblock \showarticletitle{A Formal System for Analyzing Conveyances of
  Property Under the Common Law}.
\newblock \bibinfo{journal}{\emph{JURIX}}  \bibinfo{volume}{23}
  (\bibinfo{year}{2010}), \bibinfo{pages}{139}.
\newblock


\bibitem[Beebe(2006)]%
        {beebe2006empirical}
\bibfield{author}{\bibinfo{person}{Barton Beebe}.}
  \bibinfo{year}{2006}\natexlab{}.
\newblock \showarticletitle{An empirical study of the multifactor tests for
  trademark infringement}.
\newblock \bibinfo{journal}{\emph{Calif. L. Rev.}}  \bibinfo{volume}{94}
  (\bibinfo{year}{2006}), \bibinfo{pages}{1581}.
\newblock


\bibitem[Ben-Shahar and Porat(2021)]%
        {ben2021personalized}
\bibfield{author}{\bibinfo{person}{Omri Ben-Shahar} {and}
  \bibinfo{person}{Ariel Porat}.} \bibinfo{year}{2021}\natexlab{}.
\newblock \bibinfo{booktitle}{\emph{Personalized Law: Different Rules for
  Different People}}.
\newblock \bibinfo{publisher}{Oxford University Press}.
\newblock


\bibitem[Bench-Capon et~al\mbox{.}(1987)]%
        {bench1987logic}
\bibfield{author}{\bibinfo{person}{J.M.~Trevor Bench-Capon},
  \bibinfo{person}{Gwen~O. Robinson}, \bibinfo{person}{Tom~W. Routen}, {and}
  \bibinfo{person}{Marek~J. Sergot}.} \bibinfo{year}{1987}\natexlab{}.
\newblock \showarticletitle{Logic Programming for Large Scale Applications in
  Law: A Formalisation of Supplementary Benefit Legislation}. In
  \bibinfo{booktitle}{\emph{Proc. 1st Int'l Conf. on Artificial Intelligence \&
  L.}} \bibinfo{publisher}{Association for Computing Machinery},
  \bibinfo{pages}{190}.
\newblock


\bibitem[Bergin and Haskell(1984)]%
        {berginhaskell}
\bibfield{author}{\bibinfo{person}{Thomas~F. Bergin} {and}
  \bibinfo{person}{Paul~G. Haskell}.} \bibinfo{year}{1984}\natexlab{}.
\newblock \bibinfo{booktitle}{\emph{Preface to Estates in Land and Future
  Interests} (\bibinfo{edition}{2nd ed.} ed.)}.
\newblock \bibinfo{publisher}{Foundation Press}.
\newblock


\bibitem[Blackwell(2000)]%
        {blackwell1999finally}
\bibfield{author}{\bibinfo{person}{Thomas~F. Blackwell}.}
  \bibinfo{year}{2000}\natexlab{}.
\newblock \showarticletitle{Finally Adding Method to Madness: Applying
  Principles of Object-Oriented Analysis and Design to Legislative Drafting}.
\newblock \bibinfo{journal}{\emph{NYU. J. Legis. \& Pub. Pol'y}}
  \bibinfo{volume}{3} (\bibinfo{year}{2000}), \bibinfo{pages}{227}.
\newblock


\bibitem[Borrelli(1989)]%
        {borrelli1989prolog}
\bibfield{author}{\bibinfo{person}{Marc~A Borrelli}.}
  \bibinfo{year}{1989}\natexlab{}.
\newblock \showarticletitle{Prolog and the law: Using expert systems to perform
  legal analysis in the uk}.
\newblock \bibinfo{journal}{\emph{Software LJ}}  \bibinfo{volume}{3}
  (\bibinfo{year}{1989}), \bibinfo{pages}{687}.
\newblock


\bibitem[Burk(2019)]%
        {burk2019algorithmic}
\bibfield{author}{\bibinfo{person}{Dan~L Burk}.}
  \bibinfo{year}{2019}\natexlab{}.
\newblock \showarticletitle{Algorithmic fair use}.
\newblock \bibinfo{journal}{\emph{U. Chi. L. Rev.}}  \bibinfo{volume}{86}
  (\bibinfo{year}{2019}), \bibinfo{pages}{283}.
\newblock


\bibitem[Camilleri et~al\mbox{.}(2014)]%
        {camilleri}
\bibfield{author}{\bibinfo{person}{John~J. Camilleri},
  \bibinfo{person}{Gabriele Paganelli}, {and} \bibinfo{person}{Gerardo
  Schneider}.} \bibinfo{year}{2014}\natexlab{}.
\newblock \showarticletitle{A CNL for Contract-Oriented Diagrams}.
\newblock \bibinfo{journal}{\emph{Controlled Nat. Language}}
  (\bibinfo{year}{2014}), \bibinfo{pages}{135}.
\newblock


\bibitem[Casey and Niblett(2016)]%
        {casey2016self}
\bibfield{author}{\bibinfo{person}{Anthony~J Casey} {and}
  \bibinfo{person}{Anthony Niblett}.} \bibinfo{year}{2016}\natexlab{}.
\newblock \showarticletitle{Self-driving laws}.
\newblock \bibinfo{journal}{\emph{University of Toronto Law Journal}}
  \bibinfo{volume}{66}, \bibinfo{number}{4} (\bibinfo{year}{2016}),
  \bibinfo{pages}{429--442}.
\newblock


\bibitem[Casey and Niblett(2017)]%
        {casey2017self}
\bibfield{author}{\bibinfo{person}{Anthony~J Casey} {and}
  \bibinfo{person}{Anthony Niblett}.} \bibinfo{year}{2017}\natexlab{}.
\newblock \showarticletitle{Self-driving contracts}.
\newblock \bibinfo{journal}{\emph{J. Corp. L.}}  \bibinfo{volume}{43}
  (\bibinfo{year}{2017}), \bibinfo{pages}{1}.
\newblock


\bibitem[Chesney(2021)]%
        {chesney2021cybersecurity}
\bibfield{author}{\bibinfo{person}{Robert Chesney}.}
  \bibinfo{year}{2021}\natexlab{}.
\newblock \bibinfo{booktitle}{\emph{Cybersecurity Law, Policy, and
  Institutions}}.
\newblock
\urldef\tempurl%
\url{https://papers.ssrn.com/sol3/papers.cfm?abstract_id=3547103}
\showURL{%
\tempurl}


\bibitem[Cohney and Hoffman(2020)]%
        {cohneytransactional}
\bibfield{author}{\bibinfo{person}{Shaanan Cohney} {and}
  \bibinfo{person}{David~A. Hoffman}.} \bibinfo{year}{2020}\natexlab{}.
\newblock \showarticletitle{Transactional Scripts in Contract Stacks}.
\newblock \bibinfo{journal}{\emph{Minn. L. Rev.}}  \bibinfo{volume}{105}
  (\bibinfo{year}{2020}), \bibinfo{pages}{319}.
\newblock


\bibitem[Crumlish and Malone(2009)]%
        {crumlish2009designing}
\bibfield{author}{\bibinfo{person}{Christian Crumlish} {and}
  \bibinfo{person}{Erin Malone}.} \bibinfo{year}{2009}\natexlab{}.
\newblock \bibinfo{booktitle}{\emph{Designing social interfaces: Principles,
  patterns, and practices for improving the user experience}}.
\newblock \bibinfo{publisher}{" O'Reilly Media"}.
\newblock


\bibitem[Cunningham(2006)]%
        {cunningham2006language}
\bibfield{author}{\bibinfo{person}{Lawrence~A. Cunningham}.}
  \bibinfo{year}{2006}\natexlab{}.
\newblock \showarticletitle{Language, Deals, and Standards: The Future of XML
  Contracts}.
\newblock \bibinfo{journal}{\emph{Wash. U. L. Rev.}}  \bibinfo{volume}{84}
  (\bibinfo{year}{2006}), \bibinfo{pages}{313}.
\newblock


\bibitem[Debessonet and Cross(1986)]%
        {debessonet1986artificial}
\bibfield{author}{\bibinfo{person}{Cary~G. Debessonet} {and}
  \bibinfo{person}{George~R. Cross}.} \bibinfo{year}{1986}\natexlab{}.
\newblock \showarticletitle{An Artificial Intelligence Application in the Law:
  CCLIPS, A Computer Program that Processes Legal Information}.
\newblock \bibinfo{journal}{\emph{High Tech. L.J.}}  \bibinfo{volume}{1}
  (\bibinfo{year}{1986}), \bibinfo{pages}{329}.
\newblock


\bibitem[Dung and Sartor(2011)]%
        {dung2011modular}
\bibfield{author}{\bibinfo{person}{Phan~Minh Dung} {and}
  \bibinfo{person}{Giovanni Sartor}.} \bibinfo{year}{2011}\natexlab{}.
\newblock \showarticletitle{The Modular Logic of Private International Law}.
\newblock \bibinfo{journal}{\emph{Artificial Intelligence \& L.}}
  \bibinfo{volume}{19} (\bibinfo{year}{2011}), \bibinfo{pages}{233}.
\newblock


\bibitem[Easterbrook(1996)]%
        {easterbrook1996cyberspace}
\bibfield{author}{\bibinfo{person}{Frank~H Easterbrook}.}
  \bibinfo{year}{1996}\natexlab{}.
\newblock \showarticletitle{Cyberspace and the Law of the Horse}.
\newblock \bibinfo{journal}{\emph{University of Chicago Legal Forum}}
  \bibinfo{volume}{1996} (\bibinfo{year}{1996}), \bibinfo{pages}{207}.
\newblock


\bibitem[Edwards(2009)]%
        {edwards}
\bibfield{author}{\bibinfo{person}{Linda Edwards}.}
  \bibinfo{year}{2009}\natexlab{}.
\newblock \bibinfo{booktitle}{\emph{Estates in Land and Future Interests: A
  Step-by-Step Guide} (\bibinfo{edition}{3rd ed.} ed.)}.
\newblock \bibinfo{publisher}{Wolters Kluwer}.
\newblock


\bibitem[.Finan(1981)]%
        {finan1981lawgical}
\bibfield{author}{\bibinfo{person}{John~P .Finan}.}
  \bibinfo{year}{1981}\natexlab{}.
\newblock \showarticletitle{LAWGICAL: Jurisprudential and Logical
  Considerations}.
\newblock \bibinfo{journal}{\emph{Akron L. Rev.}}  \bibinfo{volume}{15}
  (\bibinfo{year}{1981}), \bibinfo{pages}{675}.
\newblock


\bibitem[{Financial Domain-Specific Language Listing}(2022)]%
        {findsl}
\bibfield{author}{\bibinfo{person}{{Financial Domain-Specific Language
  Listing}}.} \bibinfo{year}{[2022]}\natexlab{}.
\newblock \bibinfo{booktitle}{}.
\newblock
\urldef\tempurl%
\url{http://www.dslfin.org/resources.html}
\showURL{%
\tempurl}


\bibitem[Fowler(2010)]%
        {fowler2010domain}
\bibfield{author}{\bibinfo{person}{Martin Fowler}.}
  \bibinfo{year}{2010}\natexlab{}.
\newblock \bibinfo{booktitle}{\emph{Domain-Specific Languages}}.
\newblock \bibinfo{publisher}{Addison-Wesley}.
\newblock


\bibitem[Gabriel(1996)]%
        {gabriel1996patterns}
\bibfield{author}{\bibinfo{person}{Richard~P Gabriel}.}
  \bibinfo{year}{1996}\natexlab{}.
\newblock \bibinfo{booktitle}{\emph{Patterns of Software}}.
\newblock \bibinfo{publisher}{Oxford University Press}.
\newblock


\bibitem[Gamma et~al\mbox{.}(1995)]%
        {gamma1995elements}
\bibfield{author}{\bibinfo{person}{Erich Gamma}, \bibinfo{person}{Richard
  Helm}, \bibinfo{person}{Ralph Johnson}, {and} \bibinfo{person}{John
  Vlissides}.} \bibinfo{year}{1995}\natexlab{}.
\newblock \bibinfo{booktitle}{\emph{Design Patterns Elements of Reusable
  Object-Oriented Software}}.
\newblock \bibinfo{publisher}{Addison-Wesley}.
\newblock


\bibitem[Garber(2020)]%
        {garber2020pair}
\bibfield{author}{\bibinfo{person}{Jason Garber}.}
  \bibinfo{year}{2020}\natexlab{}.
\newblock \bibinfo{booktitle}{\emph{Practical pair programming}}.
\newblock \bibinfo{publisher}{A Book Apart}.
\newblock


\bibitem[Gerding(2013)]%
        {gerding2013contract}
\bibfield{author}{\bibinfo{person}{Erik~F. Gerding}.}
  \bibinfo{year}{2013}\natexlab{}.
\newblock \showarticletitle{Contract as Pattern Language}.
\newblock \bibinfo{journal}{\emph{Wash. L. Rev.}}  \bibinfo{volume}{88}
  (\bibinfo{year}{2013}), \bibinfo{pages}{1323}.
\newblock


\bibitem[Grimmelmann(2013)]%
        {grimmelmann2013casino}
\bibfield{author}{\bibinfo{person}{James Grimmelmann}.}
  \bibinfo{year}{2013}\natexlab{}.
\newblock \bibinfo{title}{Computer Crime Law Goes to the Casino}.
\newblock
\newblock
\urldef\tempurl%
\url{https://www.techpolicy.com/Grimmelmann_ComputerCrimeLawGoesToCasino.aspx}
\showURL{%
\tempurl}


\bibitem[Grimmelmann(2016)]%
        {grimmelmann2016consenting}
\bibfield{author}{\bibinfo{person}{James Grimmelmann}.}
  \bibinfo{year}{2016}\natexlab{}.
\newblock \showarticletitle{Consenting to Computer Use}.
\newblock \bibinfo{journal}{\emph{Geo. Wash. L. Rev.}}  \bibinfo{volume}{84}
  (\bibinfo{year}{2016}), \bibinfo{pages}{1500}.
\newblock


\bibitem[Grimmelmann(2019)]%
        {grimmelmann2019all}
\bibfield{author}{\bibinfo{person}{James Grimmelmann}.}
  \bibinfo{year}{2019}\natexlab{}.
\newblock \showarticletitle{All Smart Contracts Are Ambiguous}.
\newblock \bibinfo{journal}{\emph{J. L. \& Innovation}}  \bibinfo{volume}{2}
  (\bibinfo{year}{2019}), \bibinfo{pages}{1}.
\newblock


\bibitem[Grimmelmann et~al\mbox{.}(2022)]%
        {grimmelmann2022language}
\bibfield{author}{\bibinfo{person}{James Grimmelmann},
  \bibinfo{person}{Shrutarshi Basu}, \bibinfo{person}{Nate Foster},
  \bibinfo{person}{Shan Parikh}, {and} \bibinfo{person}{Ryan Richardson}.}
  \bibinfo{year}{2022}\natexlab{}.
\newblock \showarticletitle{A Programming Language for Estates and Future
  Interests}.
\newblock \bibinfo{journal}{\emph{Yale Journal of Law and Technology}}
  \bibinfo{volume}{24} (\bibinfo{year}{2022}), \bibinfo{pages}{to appear}.
\newblock


\bibitem[Gruner(1989)]%
        {gruner1989sentencing}
\bibfield{author}{\bibinfo{person}{Richard~S. Gruner}.}
  \bibinfo{year}{1989}\natexlab{}.
\newblock \showarticletitle{Sentencing Advisor: An Expert Computer System for
  Federal Sentencing Analyses}.
\newblock \bibinfo{journal}{\emph{Santa Clara Computer \& High Tech. L.J.}}
  \bibinfo{volume}{5} (\bibinfo{year}{1989}), \bibinfo{pages}{51}.
\newblock


\bibitem[Hohfeld(1913)]%
        {hohfeld1913applied}
\bibfield{author}{\bibinfo{person}{Wesley~Newcomb Hohfeld}.}
  \bibinfo{year}{1913}\natexlab{}.
\newblock \showarticletitle{Some Fundamental Legal Conceptions as Applied in
  Judicial Legal Reasoning}.
\newblock \bibinfo{journal}{\emph{Yale L.J.}}  \bibinfo{volume}{16}
  (\bibinfo{year}{1913}), \bibinfo{pages}{28--59}.
\newblock


\bibitem[Hohfeld(1917)]%
        {hohfeld1917fundamental}
\bibfield{author}{\bibinfo{person}{Wesley~Newcomb Hohfeld}.}
  \bibinfo{year}{1917}\natexlab{}.
\newblock \showarticletitle{Fundamental legal conceptions as applied in
  judicial reasoning}.
\newblock \bibinfo{journal}{\emph{The Yale Law Journal}} \bibinfo{volume}{26},
  \bibinfo{number}{8} (\bibinfo{year}{1917}), \bibinfo{pages}{710--770}.
\newblock


\bibitem[Holzenberger et~al\mbox{.}(2020)]%
        {holzenberger2020dataset}
\bibfield{author}{\bibinfo{person}{Nils Holzenberger}, \bibinfo{person}{Andrew
  Blair-Stanek}, {and} \bibinfo{person}{Benjamin~Van Durme}.}
  \bibinfo{year}{2020}\natexlab{}.
\newblock \showarticletitle{A Dataset for Statutory Reasoning in Tax Law
  Entailment and Question Answering}. In \bibinfo{booktitle}{\emph{Proc. 2020
  Nat. Legal Language Processing (NLLP) Workshop}}.
  \bibinfo{publisher}{Association for Computational Linguistics},
  \bibinfo{pages}{31}.
\newblock


\bibitem[Hwang(2015)]%
        {hwang2015unbundled}
\bibfield{author}{\bibinfo{person}{Cathy Hwang}.}
  \bibinfo{year}{2015}\natexlab{}.
\newblock \showarticletitle{Unbundled Bargains: Multi-Agreement Dealmaking in
  Complex Mergers and Acquisitions}.
\newblock \bibinfo{journal}{\emph{U. Pa. L. Rev.}}  \bibinfo{volume}{164}
  (\bibinfo{year}{2015}), \bibinfo{pages}{1403}.
\newblock


\bibitem[Hwang and Jennejohn(2018)]%
        {hwang2018deal}
\bibfield{author}{\bibinfo{person}{Cathy Hwang} {and} \bibinfo{person}{Matthew
  Jennejohn}.} \bibinfo{year}{2018}\natexlab{}.
\newblock \showarticletitle{Deal Structure}.
\newblock \bibinfo{journal}{\emph{Nw. U. L. Rev.}}  \bibinfo{volume}{113}
  (\bibinfo{year}{2018}), \bibinfo{pages}{279}.
\newblock


\bibitem[Jones et~al\mbox{.}(2000)]%
        {jones2000composing}
\bibfield{author}{\bibinfo{person}{Simon~Peyton Jones},
  \bibinfo{person}{Jean-Marc Eber}, {and} \bibinfo{person}{Julian Seward}.}
  \bibinfo{year}{2000}\natexlab{}.
\newblock \showarticletitle{Composing contracts: an adventure in financial
  engineering (functional pearl)}.
\newblock \bibinfo{journal}{\emph{ACM SIGPLAN Notices}} \bibinfo{volume}{35},
  \bibinfo{number}{9} (\bibinfo{year}{2000}), \bibinfo{pages}{280--292}.
\newblock


\bibitem[Jupyter Project(2022)]%
        {jupyter}
Project Jupyter \bibinfo{year}{[2022]}\natexlab{}.
\newblock \bibinfo{booktitle}{\emph{{Project Jupyter}}}.
\newblock Project Jupyter.
\newblock
\urldef\tempurl%
\url{https://jupyter.org}
\showURL{%
\tempurl}


\bibitem[Kaminski(2019)]%
        {kaminski2019right}
\bibfield{author}{\bibinfo{person}{Margot~E Kaminski}.}
  \bibinfo{year}{2019}\natexlab{}.
\newblock \showarticletitle{The right to explanation, explained}.
\newblock \bibinfo{journal}{\emph{Berkeley Tech. LJ}}  \bibinfo{volume}{34}
  (\bibinfo{year}{2019}), \bibinfo{pages}{189}.
\newblock


\bibitem[Katz and Bommarito(2014)]%
        {katz2014measuring}
\bibfield{author}{\bibinfo{person}{Daniel~Martin Katz} {and}
  \bibinfo{person}{Michael~James Bommarito}.} \bibinfo{year}{2014}\natexlab{}.
\newblock \showarticletitle{Measuring the Complexity of the Law: The United
  States Code}.
\newblock \bibinfo{journal}{\emph{Artificial Intelligence \& L.}}
  \bibinfo{volume}{22}, \bibinfo{number}{4} (\bibinfo{year}{2014}),
  \bibinfo{pages}{337}.
\newblock


\bibitem[Kerr(2018)]%
        {kerr2018computer}
\bibfield{author}{\bibinfo{person}{Orin~S Kerr}.}
  \bibinfo{year}{2018}\natexlab{}.
\newblock \bibinfo{booktitle}{\emph{Computer Crime Law} (\bibinfo{edition}{4}
  ed.)}.
\newblock \bibinfo{publisher}{Thomson/West}.
\newblock


\bibitem[Klass(2022a)]%
        {klass2022vending}
\bibfield{author}{\bibinfo{person}{Gregory Klass}.}
  \bibinfo{year}{2022}\natexlab{a}.
\newblock \bibinfo{title}{How to Interpret a Vending Machine: Smart Contracts
  and Contract Law}.  (\bibinfo{year}{2022}).
\newblock
\urldef\tempurl%
\url{https://papers.ssrn.com/sol3/papers.cfm?abstract_id=4045711}
\showURL{%
\tempurl}


\bibitem[Klass(2022b)]%
        {klass2022tailoring}
\bibfield{author}{\bibinfo{person}{Gregory Klass}.}
  \bibinfo{year}{2022}\natexlab{b}.
\newblock \showarticletitle{Tailoring ex Machina: Perspectives on Personalized
  Law}.
\newblock \bibinfo{journal}{\emph{U. Chi. L. Rev. Online}}
  (\bibinfo{year}{2022}).
\newblock


\bibitem[Klemens(2005)]%
        {klemens2005math}
\bibfield{author}{\bibinfo{person}{Ben Klemens}.}
  \bibinfo{year}{2005}\natexlab{}.
\newblock \bibinfo{booktitle}{\emph{Math You Can't Use: Patents, Copyright, and
  Software}}.
\newblock \bibinfo{publisher}{Brookings Institution Press}.
\newblock


\bibitem[Knuth(1984)]%
        {knuth1984literate}
\bibfield{author}{\bibinfo{person}{Donald~Ervin Knuth}.}
  \bibinfo{year}{1984}\natexlab{}.
\newblock \showarticletitle{Literate Programming}.
\newblock \bibinfo{journal}{\emph{Computer J.}} \bibinfo{volume}{27},
  \bibinfo{number}{2} (\bibinfo{year}{1984}), \bibinfo{pages}{97}.
\newblock


\bibitem[Ladleif and Weske(2019)]%
        {ladleif2019unifying}
\bibfield{author}{\bibinfo{person}{Jan Ladleif} {and} \bibinfo{person}{Mathias
  Weske}.} \bibinfo{year}{2019}\natexlab{}.
\newblock \showarticletitle{A Unifying Model of Legal Smart Contracts}. In
  \bibinfo{booktitle}{\emph{Proc. Int'l Conf. on Conceptual Modeling}}.
  \bibinfo{pages}{323}.
\newblock


\bibitem[Landin(1966)]%
        {landin1966next}
\bibfield{author}{\bibinfo{person}{Peter~J. Landin}.}
  \bibinfo{year}{1966}\natexlab{}.
\newblock \showarticletitle{The Next 700 Programming Languages}.
\newblock \bibinfo{journal}{\emph{Comm. ACM}} \bibinfo{volume}{9},
  \bibinfo{number}{3} (\bibinfo{year}{1966}), \bibinfo{pages}{157}.
\newblock


\bibitem[Lawsky(2017a)]%
        {lawsky2017nonmonotonic}
\bibfield{author}{\bibinfo{person}{Sarah Lawsky}.}
  \bibinfo{year}{2017}\natexlab{a}.
\newblock \emph{\bibinfo{title}{Nonmonotonic Logic and Rule-Based Legal
  Reasoning}}.
\newblock \bibinfo{thesistype}{Ph.\,D. Dissertation}.
  \bibinfo{school}{University of California, Irvine}.
\newblock
\urldef\tempurl%
\url{https://escholarship.org/uc/item/59j2j45w}
\showURL{%
\tempurl}


\bibitem[Lawsky(2020)]%
        {lawsky2020form}
\bibfield{author}{\bibinfo{person}{Sarah Lawsky}.}
  \bibinfo{year}{2020}\natexlab{}.
\newblock \showarticletitle{Form as Formalization}.
\newblock \bibinfo{journal}{\emph{Ohio St. Tech. L.J.}}  \bibinfo{volume}{16}
  (\bibinfo{year}{2020}), \bibinfo{pages}{114}.
\newblock


\bibitem[Lawsky(2022)]%
        {lawskypracticeproblems}
\bibfield{author}{\bibinfo{person}{Sarah Lawsky}.}
  \bibinfo{year}{[2022]}\natexlab{}.
\newblock \bibinfo{title}{{Lawsky Practice Problems}}.
\newblock
\newblock
\urldef\tempurl%
\url{https://www.lawskypracticeproblems.org}
\showURL{%
\tempurl}


\bibitem[Lawsky(2016)]%
        {lawsky2016formalizing}
\bibfield{author}{\bibinfo{person}{Sarah~B. Lawsky}.}
  \bibinfo{year}{2016}\natexlab{}.
\newblock \showarticletitle{Formalizing the Code}.
\newblock \bibinfo{journal}{\emph{Tax L. Revieew}}  \bibinfo{volume}{70}
  (\bibinfo{year}{2016}), \bibinfo{pages}{377}.
\newblock


\bibitem[Lawsky(2017b)]%
        {lawsky2017logic}
\bibfield{author}{\bibinfo{person}{Sarah~B. Lawsky}.}
  \bibinfo{year}{2017}\natexlab{b}.
\newblock \showarticletitle{A Logic for Statutes}.
\newblock \bibinfo{journal}{\emph{Fla. Tax Rev.}}  \bibinfo{volume}{21}
  (\bibinfo{year}{2017}), \bibinfo{pages}{60}.
\newblock


\bibitem[Lessig(1999)]%
        {lessig1999code}
\bibfield{author}{\bibinfo{person}{Lawrence Lessig}.}
  \bibinfo{year}{1999}\natexlab{}.
\newblock \bibinfo{booktitle}{\emph{Code: And Other Laws of Cyberspace}}.
\newblock


\bibitem[Levine(2016)]%
        {levine2016blockchain}
\bibfield{author}{\bibinfo{person}{Matt Levine}.}
  \bibinfo{year}{2016}\natexlab{}.
\newblock \showarticletitle{Blockchain Company’s Smart Contracts Were Dumb}.
\newblock \bibinfo{journal}{\emph{Bloomberg Opinion}} (\bibinfo{year}{2016}).
\newblock
\urldef\tempurl%
\url{https://www. bloomberg.
  com/view/articles/2016-06-17/blockchaincompany-s-smart-contracts-were-dumb}
\showURL{%
\tempurl}


\bibitem[Livermore et~al\mbox{.}(2017)]%
        {livermore2017supreme}
\bibfield{author}{\bibinfo{person}{Michael~A Livermore},
  \bibinfo{person}{Allen~B Riddell}, {and} \bibinfo{person}{Daniel~N
  Rockmore}.} \bibinfo{year}{2017}\natexlab{}.
\newblock \showarticletitle{The Supreme Court and the judicial genre}.
\newblock \bibinfo{journal}{\emph{Ariz. L. Rev.}}  \bibinfo{volume}{59}
  (\bibinfo{year}{2017}), \bibinfo{pages}{837}.
\newblock


\bibitem[Livermore and Rockmore(2019)]%
        {livermore2019law}
\bibfield{author}{\bibinfo{person}{Michael~A. Livermore} {and}
  \bibinfo{person}{Daniel~N. Rockmore}.} \bibinfo{year}{2019}\natexlab{}.
\newblock \bibinfo{booktitle}{\emph{Law as Data: Computation, Text, \& the
  Future of Legal Analysis}}.
\newblock \bibinfo{publisher}{SFI Press}.
\newblock


\bibitem[MacKenzie(2001)]%
        {mackenzie2001mechanizing}
\bibfield{author}{\bibinfo{person}{Donald MacKenzie}.}
  \bibinfo{year}{2001}\natexlab{}.
\newblock \bibinfo{booktitle}{\emph{Mechanizing Proof}}.
\newblock \bibinfo{publisher}{MIT Press}.
\newblock


\bibitem[McCarty(1976)]%
        {mccarty1976reflections}
\bibfield{author}{\bibinfo{person}{L.~Thorne McCarty}.}
  \bibinfo{year}{1976}\natexlab{}.
\newblock \showarticletitle{Reflections on TAXMAN: An Experiment In Artificial
  Intelligence And Legal Reasoning}.
\newblock \bibinfo{journal}{\emph{Harv. L. Rev.}}  \bibinfo{volume}{90}
  (\bibinfo{year}{1976}), \bibinfo{pages}{837}.
\newblock


\bibitem[McCarty(1989)]%
        {mccarty1989language}
\bibfield{author}{\bibinfo{person}{L~Thorne McCarty}.}
  \bibinfo{year}{1989}\natexlab{}.
\newblock \showarticletitle{A language for legal discourse i. basic features}.
  In \bibinfo{booktitle}{\emph{Proceedings of the 2nd international conference
  on Artificial intelligence and law}}. \bibinfo{publisher}{Association for
  Computing Machinery}, \bibinfo{pages}{180--189}.
\newblock


\bibitem[McCarty(2022)]%
        {mccarty2021position}
\bibfield{author}{\bibinfo{person}{L~Thorne McCarty}.}
  \bibinfo{year}{2022}\natexlab{}.
\newblock \showarticletitle{Position Paper: LLD is All You Need}. In
  \bibinfo{booktitle}{\emph{Programming Languages and the Law (ProLaLa)}}.
  \bibinfo{publisher}{Association for Computing Machinery}.
\newblock


\bibitem[Merigoux and Huttner(2020)]%
        {merigoux2020catala}
\bibfield{author}{\bibinfo{person}{Denis Merigoux} {and} \bibinfo{person}{Liane
  Huttner}.} \bibinfo{year}{2020}\natexlab{}.
\newblock \bibinfo{title}{Catala: Moving Towards the Future of Legal Expert
  Systems}.  (\bibinfo{year}{2020}).
\newblock
\urldef\tempurl%
\url{https://hal.inria.fr/hal-02936606/document}
\showURL{%
\tempurl}


\bibitem[Merigoux et~al\mbox{.}(2021)]%
        {merigoux2020modern}
\bibfield{author}{\bibinfo{person}{Denis Merigoux},
  \bibinfo{person}{Rapha\"{e}l Monat}, {and} \bibinfo{person}{Jonathan
  Protzenko}.} \bibinfo{year}{2021}\natexlab{}.
\newblock \showarticletitle{A Modern Compiler for the French Tax Code}. In
  \bibinfo{booktitle}{\emph{Proceedings of the 30th ACM SIGPLAN International
  Conference on Compiler Construction}} (Virtual, Republic of Korea)
  \emph{(\bibinfo{series}{CC 2021})}. \bibinfo{publisher}{Association for
  Computing Machinery}, \bibinfo{address}{New York, NY, USA},
  \bibinfo{pages}{71–82}.
\newblock
\showISBNx{9781450383257}
\urldef\tempurl%
\url{https://doi.org/10.1145/3446804.3446850}
\showDOI{\tempurl}


\bibitem[Merrill and Smith(2000)]%
        {optimalstandardization}
\bibfield{author}{\bibinfo{person}{Thomas~W. Merrill} {and}
  \bibinfo{person}{Henry~E. Smith}.} \bibinfo{year}{2000}\natexlab{}.
\newblock \showarticletitle{Optimal Standarization in the Law of Property: the
  \emph {Numerus Clausus} Principle}.
\newblock \bibinfo{journal}{\emph{Yale L.J.}}  \bibinfo{volume}{110}
  (\bibinfo{year}{2000}), \bibinfo{pages}{1}.
\newblock


\bibitem[Moglen(1999)]%
        {moglen1999anarchism}
\bibfield{author}{\bibinfo{person}{Eben Moglen}.}
  \bibinfo{year}{1999}\natexlab{}.
\newblock \showarticletitle{Anarchism Triumphant: Free Software and the Death
  of Copyright}.
\newblock \bibinfo{journal}{\emph{First Monday}}  \bibinfo{volume}{4}
  (\bibinfo{year}{1999}).
\newblock
\urldef\tempurl%
\url{https://firstmonday.org/ojs/index.php/fm/article/view/684/594}
\showURL{%
\tempurl}


\bibitem[Ohm(2009)]%
        {ohm2009computer}
\bibfield{author}{\bibinfo{person}{Paul Ohm}.} \bibinfo{year}{2009}\natexlab{}.
\newblock \showarticletitle{Computer Programming and The Law: A New Research
  Agenda}.
\newblock \bibinfo{journal}{\emph{Vilanova L. Rev.}}  \bibinfo{volume}{54}
  (\bibinfo{year}{2009}), \bibinfo{pages}{117}.
\newblock


\bibitem[Powell(1949)]%
        {powell1949}
\bibfield{author}{\bibinfo{person}{Richard~R. Powell}.}
  \bibinfo{year}{1949}\natexlab{}.
\newblock \bibinfo{booktitle}{\emph{The Law of Real Property}}.
\newblock \bibinfo{publisher}{Matthew Bender \& Company}.
\newblock


\bibitem[Prakken and Sartor(2015)]%
        {prakkensartor2015}
\bibfield{author}{\bibinfo{person}{Henry Prakken} {and}
  \bibinfo{person}{Giovanni Sartor}.} \bibinfo{year}{2015}\natexlab{}.
\newblock \showarticletitle{Law and Logic: A Review From an Argumentation
  Perspective}.
\newblock \bibinfo{journal}{\emph{Artificial Intelligence}}
  \bibinfo{volume}{227} (\bibinfo{year}{2015}), \bibinfo{pages}{214}.
\newblock


\bibitem[Restatement(1936)]%
        {rest-prop1}
 \bibinfo{year}{1936}\natexlab{}.
\newblock \bibinfo{booktitle}{\emph{Restatement (First) of Property}}.
\newblock


\bibitem[Reutlinger(1994)]%
        {reutlinger1994words}
\bibfield{author}{\bibinfo{person}{Mark Reutlinger}.}
  \bibinfo{year}{1994}\natexlab{}.
\newblock \showarticletitle{When Words Fail Me: Diagramming The Rule Against
  Perpetuities}.
\newblock \bibinfo{journal}{\emph{Mo. L. Rev.}}  \bibinfo{volume}{59}
  (\bibinfo{year}{1994}), \bibinfo{pages}{157}.
\newblock


\bibitem[Rissland(1990)]%
        {rissland1990artificial}
\bibfield{author}{\bibinfo{person}{Edwina~L. Rissland}.}
  \bibinfo{year}{1990}\natexlab{}.
\newblock \showarticletitle{Artificial Intelligence and Law: Stepping Stones to
  a Model of Legal Reasoning}.
\newblock \bibinfo{journal}{\emph{Yale L.J.}} \bibinfo{volume}{99},
  \bibinfo{number}{8} (\bibinfo{year}{1990}), \bibinfo{pages}{1957}.
\newblock


\bibitem[Roach(2016)]%
        {roach2016toward}
\bibfield{author}{\bibinfo{person}{Matthew Roach}.}
  \bibinfo{year}{2016}\natexlab{}.
\newblock \showarticletitle{Toward A New Language Of Legal Drafting}.
\newblock \bibinfo{journal}{\emph{J. High Tech. L.}}  \bibinfo{volume}{17}
  (\bibinfo{year}{2016}), \bibinfo{pages}{43}.
\newblock


\bibitem[Satoh et~al\mbox{.}(2010)]%
        {satoh2010proleg}
\bibfield{author}{\bibinfo{person}{Ken Satoh}, \bibinfo{person}{Kento Asai},
  \bibinfo{person}{Takamune Kogawa}, \bibinfo{person}{Masahiro Kubota},
  \bibinfo{person}{Megumi Nakamura}, \bibinfo{person}{Yoshiaki Nishigai},
  \bibinfo{person}{Kei Shirakawa}, {and} \bibinfo{person}{Chiaki Takano}.}
  \bibinfo{year}{2010}\natexlab{}.
\newblock \showarticletitle{PROLEG: an implementation of the presupposed
  ultimate fact theory of Japanese civil code by PROLOG technology}. In
  \bibinfo{booktitle}{\emph{JSAI international symposium on artificial
  intelligence}}. \bibinfo{publisher}{Springer}, \bibinfo{pages}{153--164}.
\newblock


\bibitem[Selbst and Barocas(2018)]%
        {selbst2018intuitive}
\bibfield{author}{\bibinfo{person}{Andrew~D Selbst} {and}
  \bibinfo{person}{Solon Barocas}.} \bibinfo{year}{2018}\natexlab{}.
\newblock \showarticletitle{The intuitive appeal of explainable machines}.
\newblock \bibinfo{journal}{\emph{Fordham L. Rev.}}  \bibinfo{volume}{87}
  (\bibinfo{year}{2018}), \bibinfo{pages}{1085}.
\newblock


\bibitem[Sergot et~al\mbox{.}(1986)]%
        {sergot1986british}
\bibfield{author}{\bibinfo{person}{Marek~J. Sergot}, \bibinfo{person}{Fariba
  Sadri}, \bibinfo{person}{Robert~A. Kowalski}, \bibinfo{person}{Frank
  Kriwaczek}, \bibinfo{person}{Peter Hammond}, {and} \bibinfo{person}{H.~Terese
  Cory}.} \bibinfo{year}{1986}\natexlab{}.
\newblock \showarticletitle{The British Nationality Act as a Logic Program}.
\newblock \bibinfo{journal}{\emph{Comm. ACM}}  \bibinfo{volume}{29}
  (\bibinfo{year}{1986}), \bibinfo{pages}{370}.
\newblock


\bibitem[Shadab(2020)]%
        {shadab2020software}
\bibfield{author}{\bibinfo{person}{Houman~B. Shadab}.}
  \bibinfo{year}{2020}\natexlab{}.
\newblock \bibinfo{booktitle}{\emph{Software is Scholarship}}.
\newblock \bibinfo{type}{{T}echnical {R}eport}. \bibinfo{institution}{MIT
  Computational Law Report}.
\newblock
\urldef\tempurl%
\url{https://papers.ssrn.com/sol3/Papers.cfm?abstract_id=3632464}
\showURL{%
\tempurl}


\bibitem[Simon(1996)]%
        {simon1996sciences}
\bibfield{author}{\bibinfo{person}{Herbert~A Simon}.}
  \bibinfo{year}{1996}\natexlab{}.
\newblock \bibinfo{booktitle}{\emph{The Sciences of the Artificial}
  (\bibinfo{edition}{3rd ed.} ed.)}.
\newblock \bibinfo{publisher}{MIT Press}.
\newblock


\bibitem[Smith(2006)]%
        {smith2006modularity}
\bibfield{author}{\bibinfo{person}{Henry~E. Smith}.}
  \bibinfo{year}{2006}\natexlab{}.
\newblock \showarticletitle{ModularIty in Contracts: Boilerplate and
  Information Flow}.
\newblock \bibinfo{journal}{\emph{Mich. L. Rev.}}  \bibinfo{volume}{104}
  (\bibinfo{year}{2006}), \bibinfo{pages}{1175}.
\newblock


\bibitem[Smith(2011)]%
        {smith2011standardization}
\bibfield{author}{\bibinfo{person}{Henry~E. Smith}.}
  \bibinfo{year}{2011}\natexlab{}.
\newblock \bibinfo{booktitle}{\emph{Standardization in Property Law}}.
\newblock \bibinfo{publisher}{Edward Elgar}, \bibinfo{pages}{148}.
\newblock


\bibitem[Smith(2012)]%
        {lawofthings}
\bibfield{author}{\bibinfo{person}{Henry~E. Smith}.}
  \bibinfo{year}{2012}\natexlab{}.
\newblock \showarticletitle{Property as the Law of Things}.
\newblock \bibinfo{journal}{\emph{Harv. L. Rev.}}  \bibinfo{volume}{125}
  (\bibinfo{year}{2012}), \bibinfo{pages}{1691}.
\newblock


\bibitem[Solidity(2022)]%
        {solidity}
 \bibinfo{year}{2022}\natexlab{}.
\newblock \bibinfo{booktitle}{\emph{Solidity Documentation}
  (\bibinfo{edition}{release 0.8.12} ed.)}.
\newblock
\urldef\tempurl%
\url{https://docs.soliditylang.org}
\showURL{%
\tempurl}


\bibitem[Solum(2014)]%
        {solum2014artificial}
\bibfield{author}{\bibinfo{person}{Lawrence~B Solum}.}
  \bibinfo{year}{2014}\natexlab{}.
\newblock \showarticletitle{Artificial meaning}.
\newblock \bibinfo{journal}{\emph{Wash. L. Rev.}}  \bibinfo{volume}{89}
  (\bibinfo{year}{2014}), \bibinfo{pages}{69}.
\newblock


\bibitem[Sprankling(2017)]%
        {sprankling2012understanding}
\bibfield{author}{\bibinfo{person}{John~G. Sprankling}.}
  \bibinfo{year}{2017}\natexlab{}.
\newblock \bibinfo{booktitle}{\emph{Understanding Property Law}
  (\bibinfo{edition}{4th ed.} ed.)}.
\newblock \bibinfo{publisher}{Carolina Academic Press}.
\newblock


\bibitem[Stark(2013)]%
        {stark2013drafting}
\bibfield{author}{\bibinfo{person}{Tina~L Stark}.}
  \bibinfo{year}{2013}\natexlab{}.
\newblock \bibinfo{booktitle}{\emph{Drafting contracts: How and why lawyers do
  what they do}}.
\newblock \bibinfo{publisher}{Wolters Kluwer}.
\newblock


\bibitem[Steele(1998)]%
        {steele1998growing}
\bibfield{author}{\bibinfo{person}{Guy~L. Steele, Jr.}}
  \bibinfo{year}{1998}\natexlab{}.
\newblock \bibinfo{title}{Growing a Language}.  (\bibinfo{year}{1998}).
\newblock
\urldef\tempurl%
\url{http://www.cs.virginia.edu/~evans/cs655/readings/steele.pdf}
\showURL{%
\tempurl}


\bibitem[Strandburg(2019)]%
        {strandburg2019rulemaking}
\bibfield{author}{\bibinfo{person}{Katherine~J Strandburg}.}
  \bibinfo{year}{2019}\natexlab{}.
\newblock \showarticletitle{Rulemaking and inscrutable automated decision
  tools}.
\newblock \bibinfo{journal}{\emph{Columbia Law Review}} \bibinfo{volume}{119},
  \bibinfo{number}{7} (\bibinfo{year}{2019}), \bibinfo{pages}{1851--1886}.
\newblock


\bibitem[Surden(2012)]%
        {surden2012computable}
\bibfield{author}{\bibinfo{person}{Harry Surden}.}
  \bibinfo{year}{2012}\natexlab{}.
\newblock \showarticletitle{Computable Contracts}.
\newblock \bibinfo{journal}{\emph{U.C. Davis L. Rev.}}  \bibinfo{volume}{46}
  (\bibinfo{year}{2012}), \bibinfo{pages}{629}.
\newblock


\bibitem[Tidwell et~al\mbox{.}(2020)]%
        {tidwell2010designing}
\bibfield{author}{\bibinfo{person}{Jenifer Tidwell}, \bibinfo{person}{Charles
  Brewer}, {and} \bibinfo{person}{Aynne Valencia}.}
  \bibinfo{year}{2020}\natexlab{}.
\newblock \bibinfo{booktitle}{\emph{Designing interfaces: Patterns for
  effective interaction design} (\bibinfo{edition}{3} ed.)}.
\newblock \bibinfo{publisher}{"O'Reilly Media"}.
\newblock


\bibitem[Tien(2000)]%
        {tien2000publishing}
\bibfield{author}{\bibinfo{person}{Lee Tien}.} \bibinfo{year}{2000}\natexlab{}.
\newblock \showarticletitle{Publishing Software As A Speech Act}.
\newblock \bibinfo{journal}{\emph{Berk. Tech. L.J.}}  \bibinfo{volume}{15}
  (\bibinfo{year}{2000}), \bibinfo{pages}{629}.
\newblock


\bibitem[{United States Court of Appeals for the Second Circuit}(1992)]%
        {altai1992}
\bibfield{author}{\bibinfo{person}{{United States Court of Appeals for the
  Second Circuit}}.} \bibinfo{year}{1992}\natexlab{}.
\newblock \bibinfo{title}{Computer Associates Intern., Inc. v. Altai, Inc., 982
  F.2d 693}.
\newblock
\newblock


\bibitem[{United States Court of Appeals for the Sixth Circuit}(2000)]%
        {junger2000}
\bibfield{author}{\bibinfo{person}{{United States Court of Appeals for the
  Sixth Circuit}}.} \bibinfo{year}{2000}\natexlab{}.
\newblock \bibinfo{title}{Junger v. Daley, 209 F.3d 481}.
\newblock
\newblock


\bibitem[{United States Supreme Court}(2007)]%
        {microsoft2007}
\bibfield{author}{\bibinfo{person}{{United States Supreme Court}}.}
  \bibinfo{year}{2007}\natexlab{}.
\newblock \bibinfo{title}{Microsoft Corp. v. AT\&T Corp., 550 U.S. 437}.
\newblock
\newblock


\bibitem[{United States Supreme Court}(2014)]%
        {alice2014}
\bibfield{author}{\bibinfo{person}{{United States Supreme Court}}.}
  \bibinfo{year}{2014}\natexlab{}.
\newblock \bibinfo{title}{Alice Corp. Pty. Ltd. v. CLS Bank Intern., 573 U.S.
  208}.
\newblock
\newblock


\bibitem[{United States Supreme Court}(2021)]%
        {oracle2021}
\bibfield{author}{\bibinfo{person}{{United States Supreme Court}}.}
  \bibinfo{year}{2021}\natexlab{}.
\newblock \bibinfo{title}{Google LLC v. Oracle America, Inc., 141 S. Ct. 1183}.
\newblock
\newblock


\bibitem[van Deursen et~al\mbox{.}(2000)]%
        {dslbib}
\bibfield{author}{\bibinfo{person}{Arie van Deursen}, \bibinfo{person}{Paul
  Klint}, {and} \bibinfo{person}{Joost Visser}.} \bibinfo{year}{June
  2000}\natexlab{}.
\newblock \showarticletitle{Domain-Specific Languages: An Annotated
  Bibliography}.
\newblock \bibinfo{journal}{\emph{SIGPLAN Notices.}} (\bibinfo{year}{June
  2000}), \bibinfo{pages}{26}.
\newblock


\bibitem[van Schewick(2012)]%
        {van2012internet}
\bibfield{author}{\bibinfo{person}{Barbara van Schewick}.}
  \bibinfo{year}{2012}\natexlab{}.
\newblock \bibinfo{booktitle}{\emph{Internet Architecture and Innovation}}.
\newblock \bibinfo{publisher}{MIT Press}.
\newblock


\bibitem[Welch(1981)]%
        {welch1981lawgical}
\bibfield{author}{\bibinfo{person}{John~T. Welch}.}
  \bibinfo{year}{1981}\natexlab{}.
\newblock \showarticletitle{LAWGICAL: An Approach to Computer-Aided Legal
  Analysis}.
\newblock \bibinfo{journal}{\emph{Akron L. Rev.}}  \bibinfo{volume}{15}
  (\bibinfo{year}{1981}), \bibinfo{pages}{655}.
\newblock


\bibitem[Wendell(2007)]%
        {wendel2007treatise}
\bibfield{author}{\bibinfo{person}{Peter~T. Wendell}.}
  \bibinfo{year}{2007}\natexlab{}.
\newblock \bibinfo{booktitle}{\emph{Possessory Estates and Future Interests
  Primer} (\bibinfo{edition}{3d ed.} ed.)}.
\newblock \bibinfo{publisher}{West Academic Publishing}.
\newblock


\bibitem[Williams and Kessler(2003)]%
        {williams2003pair}
\bibfield{author}{\bibinfo{person}{Laurie Williams} {and}
  \bibinfo{person}{Robert~R Kessler}.} \bibinfo{year}{2003}\natexlab{}.
\newblock \bibinfo{booktitle}{\emph{Pair programming illuminated}}.
\newblock \bibinfo{publisher}{Addison-Wesley Professional}.
\newblock


\bibitem[Yoo(2016)]%
        {yoo2016modularity}
\bibfield{author}{\bibinfo{person}{Christopher~S. Yoo}.}
  \bibinfo{year}{2016}\natexlab{}.
\newblock \showarticletitle{Modularity Theory and Internet Regulation}.
\newblock \bibinfo{journal}{\emph{U. Ill. L. Rev.}}  \bibinfo{volume}{2016}
  (\bibinfo{year}{2016}), \bibinfo{pages}{1}.
\newblock


\end{thebibliography}

\end{document}